\documentclass[a4paper,twoside]{article}

\usepackage{amssymb}
\usepackage{graphicx}

\usepackage[caption=false,font=footnotesize]{subfig}

\usepackage{url}

\usepackage{amsmath,amsthm,amssymb}
\usepackage{euscript}

\usepackage{array}
\usepackage{multirow}
\usepackage[flushleft]{threeparttable}

\usepackage{mdwmath}
\usepackage{mdwtab}

\usepackage{qcircuit}

\PassOptionsToPackage{lowtilde}{url}
\RequirePackage[%
    colorlinks=true,pdfstartview=FitH,%
    linkcolor=blue,citecolor=blue,filecolor=blue,urlcolor=blue,linktoc=page]{hyperref}

\newcommand{\comment}[1]{}
\newcommand{\compose}{\mathop{*}\limits}

\newcommand{\BB}{\mathbb B}
\newcommand{\frS}{\mathfrak S}
\newcommand{\vv}{\mathbf}

\theoremstyle{plain}

\newtheorem{lemma}{Lemma}

\theoremstyle{definition}
\newtheorem{definition}{Definition}

\title{\texorpdfstring{Application of Permutation Group Theory\\in Reversible Logic Synthesis}
{Application of Permutation Group Theory in Reversible Logic Synthesis}}

\author{
    {\small
        \begin{tabular}[t]{c}
        \normalsize Dmitry V. Zakablukov\medskip\\
        Dep. of Information Security, Bauman Moscow State Technical University,\\
        Moscow, Russian Federation\smallskip\\
        E-mail: \texttt{\href{mailto:dmitriy.zakablukov@gmail.com}{dmitriy.zakablukov@gmail.com}}
        \end{tabular}
    }
}

\begin{document}

\maketitle

\begin{abstract}
    The paper discusses various applications of permutation group theory in the synthesis of reversible logic circuits
    consisting of Toffoli gates with negative control lines. An asymptotically optimal synthesis algorithm
    for circuits consisting of gates from the NCT library is described.
    An algorithm for gate complexity reduction, based on equivalent replacements of gates compositions, is introduced.
    A new approach for combining a group-theory-based synthesis algorithm with a Reed--Muller-spectra-based synthesis algorithm
    is described.
    Experimental results are presented to show that the proposed synthesis techniques allow a reduction in input lines
    count, gate complexity or quantum cost of reversible circuits for various benchmark functions.
\end{abstract}


\section{Introduction}

Reversible logic circuits have been studied in many recent papers%
~\cite{abdessaied_reducing_depth}, \cite{miller_reducing_complexity},
\cite{markov_survey}, \cite{perkowski_linear}, \cite{my_gate_complexity}.
On the one hand, the interest in these circuits is caused by the theoretically possible reduction of energy consumption
in digital devices due to the reversibility of all computations~\cite{bennett}.
On the other hand, all quantum computations are necessarily reversible. Hence, with the help of a reversible circuit,
one can model a quantum circuit.

One important research area is the development of new efficient and fast synthesis algorithms,
which can produce a reversible circuit with low gate complexity and depth.
However, for the purpose of a comparison between different synthesis algorithms, we should first choose a library of gates,
from which a synthesized circuit will consist.
One such gate library is one that includes NOT (inversion gate), CNOT (Feynman gate)
and C\textsuperscript{2}NOT (Toffoli gate). We will refer to it as the NCT library.
Another popular gate library is the GT library, which includes generalized Toffoli gates
with positive and negative control input lines.
Both libraries are functionally complete in terms of the ability to construct a reversible circuit that implements
a desired even permutation from the alternating group $A(\BB^n)$ without using additional inputs.
An odd permutation from the symmetric group $S(\BB_2^n)$ can always be realized in a reversible circuit
without additional inputs in the GT, but not in the NCT library.

For many proposed synthesis algorithms, an upper bound for the gate complexity of a reversible circuit in the worst case
is proved. Though it was proved that the worst case requires $\mathrm\Omega(n2^n \mathop / \log n)$ gates
from the NCT library~\cite{shende}, almost all these bounds are of the form $\mathrm O(n2^n)$
in the NCT library~\cite{maslov_rm_synthesis}.

Recently, the first asymptotically optimal in NCT library synthesis algorithm was introduced with the gate complexity
$L(\frS) \lesssim 3n 2^{n+4} \mathop / \log_2 n$ of a reversible circuit in the worst case~\cite{my_gate_complexity}.
In Section~\ref{section_asymp_alg}, we briefly describe this cycle-based algorithm.
Section~\ref{section_rules} contains descriptions of the replacement rules from~\cite{my_equivalent_replacements}
and of a ``moving and replacing'' algorithm for reducing the gate complexity
of a reversible circuit in NCT and GT libraries with the help of these rules.
In Section~\ref{section_cube}, we discuss various approaches of reducing the gate complexity during the synthesis process.
In Section~\ref{section_rm_combination}, we introduce a novel technique for combining a cycle-based synthesis algorithm with
a Reed--Muller-spectra-based one.
Experimental results of benchmark functions synthesis are presented in Section~\ref{section_exp};
all new circuits were obtained with the help of our open source
software \textit{ReversibleLogicGenerator}~\cite{my_software} that implements all synthesis techniques described in this paper.
All results we present here (except Section~\ref{section_asymp_alg} and the first part
of Section~\ref{section_rules}) are new.

We use the following notation for a generalized Toffoli gate with negative control input lines.
\begin{definition}
    A generalized Toffoli gate $TOF(I;J;t) = TOF(i_1, \cdots, i_r; j_1, \cdots $ $\cdots, j_s;t)$ is a reversible gate,
    which defines a transformation $f_{I;J;t}\colon \BB^n \to \BB^n$ as follows:
    $$
        f_{I;J;t}(\langle x_1, \cdots, x_n \rangle) = \langle x_1, \cdots,
            x_t \oplus x_{i_1}\wedge \cdots \wedge x_{i_r}
            \wedge \bar x_{j_1} \wedge \cdots \wedge \bar x_{j_s}, \cdots, x_n \rangle \;  ,
    $$
    where $I = \{\,i_1, \cdots, i_r \,\}$ is a set of indices of positive control input lines,
    $J = \{\,j_1, \cdots, j_s \,\}$ is a set of indices of negative control input lines,
    and $t$ is an index of a controlled output line, $I \cap J = \emptyset$, $t \notin I \cup J$.
\end{definition}

In the case of the absence of negative control input lines, a generalized Toffoli gate will be referenced
as $TOF(I;t)$, and in the case when a generalized Toffoli gate has no control input lines at all,
it will be referenced as $TOF(t)$.
In other words, $TOF(t) = TOF(\emptyset;\emptyset;t)$ and $TOF(I;t) = TOF(I;\emptyset;t)$.
Using this notation, we can refer to a NOT gate as $TOF(a)$, to a CNOT gate as $TOF(b;a)$
and to a C\textsuperscript{2}NOT gate as $TOF(b,c;a)$.

\section{Asymptotically optimal synthesis algorithm}\label{section_asymp_alg}

In~\cite{my_gate_complexity} a cycle-based synthesis algorithm
that can produce a reversible circuit with the asymptotically optimal in NCT library gate complexity
for any even permutation on the set $\BB^n$, was described.
It is the first and currently (as far as we know) the only asymptotically optimal non-search synthesis algorithm
for the NCT library.
Our software~\cite{my_software} is based on it, so we are going to briefly describe the essence of the algorithm.

Let's consider an even permutation $h \in A(\BB^n)$. The main idea is a decomposition of $h$ into a product
of transpositions in such a way that all of them can be grouped by $K$ independent transpositions%
\footnote{Hereinafter a multiplication of permutations is left-associative: $(f \circ g)(x) = g(f(x))$.}:
\begin{equation*}
    h = G_1 \circ G_2 \circ \cdots \circ G_t \circ h' \;  ,
\end{equation*}
where $G_i = (\vv x_{i,1}, \vv y_{i,1}) \circ \cdots \circ (\vv x_{i,K}, \vv y_{i,K})$ is an $i$-th group
of $K$ independent transpositions, $\vv x_{i,j}, \vv y_{i,j} \in \BB^n$ and $h'$ is a residual permutation.

Using vectors of a group $G_i$, we construct a matrix $A_i$ as follows:
$$
    A_i =
            \begin{bmatrix}
                \vv x_{i,1} &
                \vv y_{i,1} &
                \cdots &
                \vv x_{i,K} &
                \vv y_{i,K}
            \end{bmatrix}^{T}  \; .
$$
The matrix $A_i$ is a $2K \times n$ binary matrix. If $2^{2K} < n$, then some columns in it are equal
to one another. These duplicated columns can be zeroed-out in the matrix, using CNOT gates, with the help of conjugation;
this results in a new matrix $A_i^{(1)}$.

Note that the matrix $A_i$ defines a permutation $\pi_i \in S(\BB^n)$ and every gate $e$ from the NCT library defines
a permutation $h_e \in S(\BB^n)$, for which $h^{-1}_e = h_e$. Therefore, a conjugation of a permutation $\pi$ by
a permutation $h_e$, denoted as $\pi^{h_e} = h^{-1}_e \circ \pi \circ h_e$, corresponds to attaching the gate $e$
to the front and back of a current sub-circuit. For example, if the first two columns in the matrix $A_i$ are equal,
we can zero-out the second column with the help of two $TOF(1;2)$ gates. 

Next, we fix all pairwise distinct nonzero columns $\{\, c_{j_1}, \cdots, c_{j_d}\,\}$
in the matrix $A_i^{(1)}$ and choose an index of a controlled output $t$ from the set $\{\, j_1, \cdots, j_d\,\}$.
After that we transform the matrix $A_i^{(1)}$ to the \textit{canonical} form $A_i^{(2)}$ with the help of conjugation,
where an $l$-th row, $l$ is odd, differs from the $(l+1)$-th row only in $t$-th element.

And finally, we transform the matrix $A_i^{(2)}$ to the final form $A_i^{(3)}$ with the help of $TOF(j)$ gates,
where $j \notin \{\, j_1, \cdots, j_d\,\}$.
In~\cite{my_gate_complexity} it was proved that the matrix $A_i^{(3)}$ can be realized by the single gate
$TOF(\{\, 1, \cdots, n\,\} \setminus \{\, j_1, \cdots, j_d\,\}; t)$.
The gate can be represented as a composition of C\textsuperscript{2}NOT gates if $K > 1$
(the number of independent transpositions in a group $G_i$).

A synthesized reversible circuit $\frS$, produced by the algorithm, has the gate complexity
$L(\frS) \lesssim 3n 2^{n+4} \mathop / \log_2 n$, if $K = O(\log_2 n - \log_2 \log_2 n - \log_2 \phi(n))$,
where $\phi(n) < n \mathop / \log_2 n$ is an arbitrarily slowly growing function,
and the gate complexity $L(\frS) \lesssim 6n2^n$, if $K = 2$. These results were proved in~\cite{my_gate_complexity}.

In our software~\cite{my_software}, we can change the parameter $K$ to achieve the best synthesis result in a particular case.
But in practice, when the number of input lines in a circuit is large,
it is almost always the best option to use $K = [\log_2 n]$ during the synthesis process. 

The time complexity of the synthesis algorithm is $T(A) = O(n2^n \mathop / \log_2 n)$ in the worst case.
\section{Generalized replacement rules for gate compositions}\label{section_rules}

One of the most widely used gate complexity reduction techniques is
an applying gate compositions templates to a reversible circuit.
For example, such templates were considered in~\cite{maslov_rm_synthesis}.
This approach involves storing templates and finding them in a circuit.
But we can interchange some adjacent gates of NCT and GT libraries in a reversible circuit
without changing the resulting transformation, defined by the circuit. We call such gates \textit{independent}.

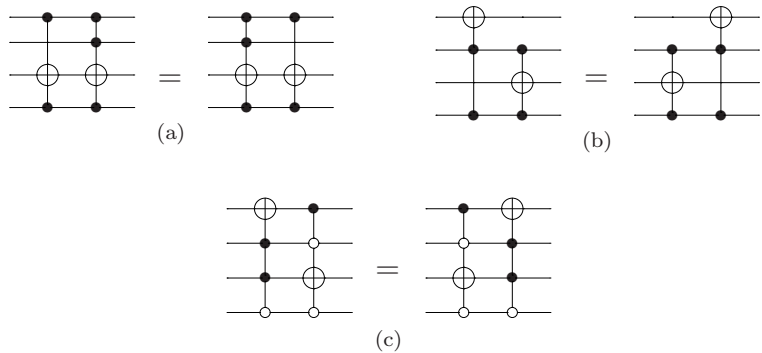
\begin{figure}
    \centering
    \begin{tabular}{cc}
        \subfloat[]{
            \Qcircuit @C=1em @R=.7em {
                & \ctrl{2}  & \ctrl{1}  & \qw \\
                & \qw       & \ctrl{1}  & \qw \\
                & \targ     & \targ     & \qw \\
                & \ctrl{-1} & \ctrl{-1} & \qw
            }
            \raisebox{-2.6em}{\hspace{2mm} \large{=} \hspace{0.9mm}}
            \Qcircuit @C=1em @R=.7em {
                & \ctrl{1}  & \ctrl{2}  & \qw \\
                & \ctrl{1}  & \qw       & \qw \\
                & \targ     & \targ     & \qw \\
                & \ctrl{-1} & \ctrl{-1} & \qw
            }\label{fig_independent_1}
        }\qquad &
        \qquad\subfloat[]{
            \Qcircuit @C=1em @R=.7em {
                & \targ     & \qw       & \qw \\
                & \ctrl{-1} & \ctrl{1}  & \qw \\
                & \qw       & \targ     & \qw \\
                & \ctrl{-2} & \ctrl{-1} & \qw
            }
            \raisebox{-2.6em}{\hspace{2mm} \large{=} \hspace{0.9mm}}
            \Qcircuit @C=1em @R=.7em {
                & \qw       & \targ     & \qw \\
                & \ctrl{1}  & \ctrl{-1} & \qw \\
                & \targ     & \qw       & \qw \\
                & \ctrl{-1} & \ctrl{-2} & \qw
            }\label{fig_independent_2}
        } \\
        & \\
        \multicolumn{2}{c}{
            \subfloat[]{
                \Qcircuit @C=1em @R=.7em {
                    & \targ      & \ctrl{1}   & \qw \\
                    & \ctrl{-1}  & \ctrlo{1}  & \qw \\
                    & \ctrl{-1}  & \targ      & \qw \\
                    & \ctrlo{-1} & \ctrlo{-1} & \qw
                }
                \raisebox{-2.6em}{\hspace{2mm} \large{=} \hspace{0.9mm}}
                \Qcircuit @C=1em @R=.7em {
                    & \ctrl{1}   & \targ      & \qw \\
                    & \ctrlo{1}  & \ctrl{-1}  & \qw \\
                    & \targ      & \ctrl{-1}  & \qw \\
                    & \ctrlo{-1} & \ctrlo{-1} & \qw
                }\label{fig_independent_3}
            }
        }
    \end{tabular}
    \caption{Examples of independent gates: (a)--(b) NCT library specific; (c) GT library specific.}
    \label{fig_independent_gates}
\end{figure}

In~\cite{iwama_transform_rules} the necessary and sufficient conditions for the independence of two $TOF(I_j;t_j)$
gates were proved.
However, for the gates from the GT library we can supplement these conditions.
\begin{lemma}\label{lemma_gate_independency}
    Gates $TOF(I_1;J_1;t_1)$ and $TOF(I_2;J_2;t_2)$ are independent iff at least one of the following condition holds
    (see Fig.~\ref{fig_independent_gates}):
    \begin{enumerate}
        \item $t_1 \notin I_2 \cup J_2$ and $t_2 \notin I_1 \cup J_1$ (in particular, $t_1 = t_2$);
        \item $I_1 \cap J_2 \ne \emptyset$ or $I_2 \cap J_1 \ne \emptyset$.
    \end{enumerate}
\end{lemma}
Proof of the Lemma~\ref{lemma_gate_independency} was partly given in~\cite{my_equivalent_replacements}.
Even though the first condition of gate independence was already known before~\cite{iwama_transform_rules}
(see Fig.~\ref{fig_independent_1}--\ref{fig_independent_2}),
the second one cannot be derived from it (see Fig.~\ref{fig_independent_3}).

In~\cite{saeedi_rule_based} rule-based optimization techniques based on Karnaugh maps for the optimization of sub-circuits
with common targets were described. The main disadvantage of this approach is the restricted scalability for circuits with
the large number of input lines. On the other hand, the advantage of using negative control Toffoli gates for the
simplification of reversible circuits and reducing their quantum cost was shown by the authors.

In~\cite{my_equivalent_replacements} we proposed generalized replacement rules
for the case of an arbitrary number of input lines.
Moreover, we were able to obtain a new rule for interchanging two gates
with changing the polarity of a control line for one of these gates
(see the last rule in Table~\ref{table_interchanging_rules}).
Our replacement rules are essentially templates of small length. But the advantage of using them is
in changing the set of negative control input lines in a gate. This makes it possible to obtain independent gates instead of
dependent ones in some cases, interchange or move them in a reversible circuit to new places and apply other replacement rules.
``Moving and replacing'' algorithm will be described later.

A similar approach was used in~\cite{rahman_rules}, though replacement rules in that paper differ from ours.

Let's consider a composition of two dependent gates $e_1 * e_2$. Let $h_{e_1}$ and $h_{e_2}$ be permutations defined by them
respectively. If we want to obtain an equal composition $\frS_1 * e_1 = e_1 * e_2$, then the circuit $\frS_1$ must implement
the permutation $h_1 = h_{e_2}^{h_{e_1}}$.
And if we want to obtain an equal composition $e_2 * \frS_2 = e_1 * e_2$, then the circuit $\frS_2$ must implement
the permutation $h_2 = h_{e_1}^{h_{e_2}}$.

All our replacement rules can be classified as follow:
\begin{enumerate}
    \item Representing a gate from the GT library as a composition of gates from the NCT library.
    \item Merging two gates into one.
    \item Reducing the negative control lines number.
    \item Interchanging two dependent gates.
\end{enumerate}

For the sake of clarity, we will organize the detailed description of our rules in the form of
Tables~\ref{table_replacing_rules}--\ref{table_interchanging_rules},
one for each rule ``class''. The left column of the tables contains a gate composition before applying replacement rule,
and the right column contains the result of the replacement. For every rule a picture goes first
(for understanding the concept of a rule), then, a text description of the rule, and finally, a condition for applying the rule.

Now we can describe the ``moving and replacing'' algorithm implemented
in our software~\cite{my_software}, which may reduce the gate complexity of a circuit.

Let a reversible circuit $\frS$ be a composition of $l$ gates from the GT library: $\frS = \compose_{i = 1}^l {e_i}$.
If a gate composition $e_i * e_j$ satisfies the condition of a replacement, where $i < j$, and there is such an index $s$,
$i \leq s < j$, that gates $e_i$ and $e_k$ are independent for every $i < k \leq s$,
and gates $e_j$ and $e_k$ are independent for every $s < k < j$, then the gates $e_i$ and $e_j$ can be removed from the
circuit and a result of the replacement for $e_i * e_j$ can be inserted between gates $e_s$ and $e_{s+1}$.

\begin{table}[t]
    \caption{Representing a gate from the GT library as a composition of gates from the NCT library.}
    \label{table_replacing_rules}
    \centering
    \begin{tabular}{|c|c|}
        \hline
        
        \vbox{\vspace{1em}\hbox to 0.2\textwidth{\hfil
            \Qcircuit @C=1em @R=.7em {
                & \ctrlo{1} & \qw \\
                & \ctrl{1}  & \qw \\
                & \ctrlo{1} & \qw \\
                & \targ     & \qw
            }
        \hfil}\vspace{0em}}
        &
        \vbox{\vspace{1em}\hbox to 0.7\textwidth{\hfil
            \Qcircuit @C=1em @R=.7em {
                & \targ & \ctrl{1} & \targ & \qw \\
                & \qw   & \ctrl{1} & \qw   & \qw \\
                & \targ & \ctrl{1} & \targ & \qw \\
                & \qw   & \targ    & \qw   & \qw
            }
        \hfil}\vspace{0em}} \\
        
        \vbox{\vspace{0em}\hbox to 0.2\textwidth{\hfil
            $TOF(I;J;t)$
        \hfil}\vspace{1.2em}}
        &
        \vbox{\vspace{0.5em}\hbox to 0.7\textwidth{\hfil
            $\left(\compose_{t\colon t \in J}{TOF(t)} \right) * TOF(I \cup J;t) *
                \left(\compose_{t\colon t \in J}{TOF(t)} \right)$
        \hfil}\vspace{0.5em}} \\
        
        \hline
        \hline

        \vbox{\vspace{1em}\hbox to 0.2\textwidth{\hfil
            \Qcircuit @C=1em @R=.7em {
                & \ctrlo{1} & \qw \\
                & \ctrl{1}  & \qw \\
                & \ctrlo{1} & \qw \\
                & \targ     & \qw
            }
        \hfil}\vspace{1em}}
        &
        \vbox{\vspace{1em}\hbox to 0.6\textwidth{\hfil
            \Qcircuit @C=1em @R=.7em {
                & \qw      & \ctrl{1} & \qw      & \ctrl{1} & \qw \\
                & \ctrl{2} & \ctrl{2} & \ctrl{1} & \ctrl{1} & \qw \\
                & \qw      & \qw      & \ctrl{1} & \ctrl{1} & \qw \\
                & \targ    & \targ    & \targ    & \targ    & \qw
            }
        \hfil}\vspace{1em}} \\
        
        
        \vbox{\vspace{0em}\hbox to 0.2\textwidth{\hfil
            $TOF(I;J;t)$
        \hfil}\vspace{0.7em}}
        &
        \vbox{\vspace{0em}\hbox to 0.6\textwidth{\hfil
            $\compose_{J'\colon J' \subseteq J}{TOF(I \cup J';t)}$
        \hfil}\vspace{0.1em}} \\
        
        \hline
    \end{tabular}
\end{table}

\begin{table}[t]
    \caption{Merging two gates into one.}
    \label{table_merging_rules}
    \centering
    \begin{tabular}{|c|c|}
        \hline
        
        \vbox{\vspace{1em}\hbox to 0.5\textwidth{\hfil
            \Qcircuit @C=1em @R=.7em {
                & \ctrl{1}  & \ctrl{1}  & \qw \\
                & \ctrlo{1} & \ctrlo{1} & \qw \\
                & \ctrl{1}  & \ctrl{1}  & \qw \\
                & \ctrl{1}  & \ctrlo{1} & \qw \\
                & \targ     & \targ     & \qw
            }
        \hfil}\vspace{0em}}
        &
        \vbox{\vspace{1em}\hbox to 0.3\textwidth{\hfil
            \Qcircuit @C=1em @R=.7em {
                & \ctrl{1}  & \qw \\
                & \ctrlo{1} & \qw \\
                & \ctrl{2}  & \qw \\
                & \qw       & \qw \\
                & \targ     & \qw
            }
        \hfil}\vspace{0em}} \\
        
        \vbox{\vspace{1em}\hbox to 0.5\textwidth{\hfil
            $TOF(I_1;J_1;t) * TOF(I_2;J_2;t)$
        \hfil}\vspace{0.5em}}
        &
        \vbox{\vspace{1em}\hbox to 0.3\textwidth{\hfil
            $TOF(I_2;J_1;t)$
        \hfil}\vspace{0.5em}} \\

        \hline
        \multicolumn{2}{|c|}{\vbox{\vspace{0.4em}\hbox{
            Condition: $I_1 = I_2 \cup \{\,k\,\}$ and $J_2 = J_1 \cup \{\,k\,\}$, where $k \notin I_2 \cup J_1$.
        }\vspace{0.05em}}} \\
        
        \hline
        \hline

        \vbox{\vspace{1em}\hbox to 0.5\textwidth{\hfil
            \Qcircuit @C=1em @R=.7em {
                & \ctrl{1}  & \ctrl{1}  & \qw \\
                & \ctrlo{1} & \ctrlo{1} & \qw \\
                & \ctrl{1}  & \ctrl{2}  & \qw \\
                & \ctrl{1}  & \qw       & \qw \\
                & \targ     & \targ     & \qw
            }
        \hfil}\vspace{0em}}
        &
        \vbox{\vspace{1em}\hbox to 0.3\textwidth{\hfil
            \Qcircuit @C=1em @R=.7em {
                & \ctrl{1}  & \qw \\
                & \ctrlo{1} & \qw \\
                & \ctrl{1}  & \qw \\
                & \ctrlo{1} & \qw \\
                & \targ     & \qw
            }
        \hfil}\vspace{0em}} \\
        
        \vbox{\vspace{1em}\hbox to 0.5\textwidth{\hfil
            $TOF(I_1;J;t) * TOF(I_2;J;t)$
        \hfil}\vspace{0.5em}}
        &
        \vbox{\vspace{1em}\hbox to 0.3\textwidth{\hfil
            $TOF(I_2; J \cup \{\,k\,\};t)$
        \hfil}\vspace{0.5em}} \\

        \hline
        \multicolumn{2}{|c|}{\vbox{\vspace{0.4em}\hbox{
            Condition: $I_1 = I_2 \cup \{\,k\,\}$.
        }\vspace{0.05em}}} \\
        
        \hline
        \hline

        \vbox{\vspace{1em}\hbox to 0.5\textwidth{\hfil
            \Qcircuit @C=1em @R=.7em {
                & \ctrl{1}  & \ctrl{1}  & \qw \\
                & \ctrlo{1} & \ctrlo{1} & \qw \\
                & \ctrl{1}  & \ctrl{2}  & \qw \\
                & \ctrlo{1} & \qw       & \qw \\
                & \targ     & \targ     & \qw
            }
        \hfil}\vspace{0em}}
        &
        \vbox{\vspace{1em}\hbox to 0.3\textwidth{\hfil
            \Qcircuit @C=1em @R=.7em {
                & \ctrl{1}  & \qw \\
                & \ctrlo{1} & \qw \\
                & \ctrl{1}  & \qw \\
                & \ctrl{1}  & \qw \\
                & \targ     & \qw
            }
        \hfil}\vspace{0em}} \\
        
        \vbox{\vspace{1em}\hbox to 0.5\textwidth{\hfil
            $TOF(I;J_1;t) * TOF(I;J_2;t)$
        \hfil}\vspace{0.5em}}
        &
        \vbox{\vspace{1em}\hbox to 0.3\textwidth{\hfil
            $TOF(I \cup \{\,k\,\}; J_2;t)$
        \hfil}\vspace{0.5em}} \\

        \hline
        \multicolumn{2}{|c|}{\vbox{\vspace{0.4em}\hbox{
            Condition: $J_1 = J_2 \cup \{\,k\,\}$.
        }\vspace{0.05em}}} \\
        
        \hline
    \end{tabular}
\end{table}

\begin{table}[t]
    \caption{Reducing the negative control lines number.}
    \label{table_reducing_rules}
    \centering
    \begin{tabular}{|c|c|}
        \hline
        
        \vbox{\vspace{1em}\hbox to 0.4\textwidth{\hfil
            \Qcircuit @C=1em @R=.7em {
                & \ctrlo{1} & \ctrlo{1} & \qw \\
                & \ctrl{1}  & \ctrl{1}  & \qw \\
                & \ctrlo{1} & \ctrl{1}  & \qw \\
                & \ctrl{1}  & \ctrlo{1} & \qw \\
                & \targ     & \targ     & \qw
            }
        \hfil}\vspace{0em}}
        &
        \vbox{\vspace{1em}\hbox to 0.5\textwidth{\hfil
            \Qcircuit @C=1em @R=.7em {
                & \ctrlo{1} & \ctrlo{1} & \qw \\
                & \ctrl{2}  & \ctrl{1}  & \qw \\
                & \qw       & \ctrl{2}  & \qw \\
                & \ctrl{1}  & \qw       & \qw \\
                & \targ     & \targ     & \qw
            }
        \hfil}\vspace{0em}} \\
        
        \vbox{\vspace{1em}\hbox to 0.4\textwidth{\hfil
            $TOF(I_1;J_1;t) * TOF(I_2;J_2;t)$
        \hfil}\vspace{1.2em}}
        &
        \vbox{\vspace{1em}\hbox to 0.5\textwidth{\begin{minipage}{0.5\textwidth}\begin{center}
            $TOF(I_1;J_3;t) * TOF(I_2;J_3;t)$ \\
            $J_3 = J_1 \setminus \{\,q\,\} = J_2 \setminus \{\,p\,\}$
        \end{center}\end{minipage}}\vspace{0.5em}} \\
        
        \hline
        \multicolumn{2}{|c|}{\vbox{\vspace{0.4em}\hbox{\begin{minipage}{0.9\textwidth}\begin{center}
            Condition: there are such $p$ and $q$, that $p \in I_1 \cap J_2$, $q \in J_1 \cap I_2$, \\
            $I_2 = I_1 \setminus \{\,p\,\} \cup \{\,q\,\}$, $J_2 = J_1 \setminus \{\,q\,\} \cup \{\,p\,\}$.
        \end{center}\end{minipage}}\vspace{0.05em}}} \\
        
        \hline
        \hline
        
        \vbox{\vspace{1em}\hbox to 0.4\textwidth{\hfil
            \Qcircuit @C=1em @R=.7em {
                & \ctrlo{1} & \ctrlo{1} & \qw \\
                & \ctrl{1}  & \ctrl{2}  & \qw \\
                & \ctrlo{2} & \qw       & \qw \\
                & \qw       & \ctrlo{1} & \qw \\
                & \targ     & \targ     & \qw
            }
        \hfil}\vspace{0em}}
        &
        \vbox{\vspace{1em}\hbox to 0.5\textwidth{\hfil
            \Qcircuit @C=1em @R=.7em {
                & \ctrlo{1} & \ctrlo{1} & \qw \\
                & \ctrl{1}  & \ctrl{2}  & \qw \\
                & \ctrl{2}  & \qw       & \qw \\
                & \qw       & \ctrl{1}  & \qw \\
                & \targ     & \targ     & \qw
            }
        \hfil}\vspace{0em}} \\
        
        \vbox{\vspace{1em}\hbox to 0.4\textwidth{\hfil
            $TOF(I_1;J_1;t) * TOF(I_2;J_2;t)$
        \hfil}\vspace{1.2em}}
        &
        \vbox{\vspace{1em}\hbox to 0.5\textwidth{\begin{minipage}{0.5\textwidth}\begin{center}
            $TOF(I_1 \cup \{\,p\,\};J_1 \setminus \{\,p\,\};t) * $ \\
            $* TOF(I_2 \cup \{\,q\,\};J_2 \setminus \{\,q\,\};t)$
        \end{center}\end{minipage}}\vspace{0.5em}} \\
        
        \hline
        \multicolumn{2}{|c|}{\vbox{\vspace{0.4em}\hbox{\begin{minipage}{0.9\textwidth}\begin{center}
            Condition: there are such $p$ and $q$, that $J_1 = J' \cup \{\,p\,\}$, $J_2 = J' \cup \{\,q\,\}$, \\
            $J_1 \cap J_2 = J'$, $I_1 = I_2$.
        \end{center}\end{minipage}}\vspace{0.05em}}} \\
        \hline
    \end{tabular}
\end{table}

\begin{table}[t]
    \caption{Interchanging two dependent gates.}
    \label{table_interchanging_rules}
    \centering
    \begin{tabular}{|c|c|}
        \hline
        
        \vbox{\vspace{1em}\hbox to 0.4\textwidth{\hfil
            \Qcircuit @C=1em @R=.7em {
                & \qw        & \ctrl{1}   & \qw \\
                & \targ      & \ctrlo{2}  & \qw \\
                & \ctrlo{-1} & \qw        & \qw \\
                & \ctrl{-1}  & \ctrl{1}   & \qw \\
                & \qw        & \targ      & \qw
            }
        \hfil}\vspace{0em}}
        &
        \vbox{\vspace{1em}\hbox to 0.5\textwidth{\hfil
            \Qcircuit @C=1em @R=.7em {
                & \ctrl{2}   & \ctrl{1}   & \qw        & \qw \\
                & \qw        & \ctrlo{2}  & \targ      & \qw \\
                & \ctrlo{1}  & \qw        & \ctrlo{-1} & \qw \\
                & \ctrl{1}   & \ctrl{1}   & \ctrl{-1}  & \qw \\
                & \targ      & \targ      & \qw        & \qw
            }
        \hfil}\vspace{0em}} \\
        
        \vbox{\vspace{1em}\hbox to 0.4\textwidth{\hfil
            $TOF(I_1;J_1;t_1) * TOF(I_2;J_2;t_2)$
        \hfil}\vspace{1.8em}}
        &
        \vbox{\vspace{1em}\hbox to 0.5\textwidth{\begin{minipage}{0.5\textwidth}\begin{center}
            $TOF(I_3;J_3;t_2) * TOF(I_2;J_2;t_2) *$ \\
            $* TOF(I_1;J_1;t_1)$ \\
            $I_3 = I_1 \cup I_2 \setminus \{\,t_1\,\}$, $J_3 = J_1 \cup J_2 \setminus \{\,t_1\,\}$
        \end{center}\end{minipage}}\vspace{0.5em}} \\
        
        \hline
        \multicolumn{2}{|c|}{\vbox{\vspace{0.4em}\hbox{
            Condition: gates are dependent, $t_1 \in I_2 \cup J_2$ and $t_2 \notin I_1 \cup J_1$.
        }\vspace{0.05em}}} \\

        \hline
        \hline

        \vbox{\vspace{1em}\hbox to 0.4\textwidth{\hfil
            \Qcircuit @C=1em @R=.7em {
                & \ctrlo{2}  & \ctrlo{1} & \qw \\
                & \qw        & \ctrl{1}  & \qw \\
                & \targ      & \ctrlo{1} & \qw \\
                & \qw        & \ctrlo{1} & \qw \\
                & \ctrl{-2}  & \ctrl{1}  & \qw \\
                & \qw        & \targ     & \qw
            }
        \hfil}\vspace{0em}}
        &
        \vbox{\vspace{1em}\hbox to 0.5\textwidth{\hfil
            \Qcircuit @C=1em @R=.7em {
                & \ctrlo{1} & \ctrlo{2}  & \qw \\
                & \ctrl{1}  & \qw        & \qw \\
                & \ctrl{1}  & \targ      & \qw \\
                & \ctrlo{1} & \qw        & \qw \\
                & \ctrl{1}  & \ctrl{-2}  & \qw \\
                & \targ     & \qw        & \qw
            }
        \hfil}\vspace{0em}} \\
        
        \vbox{\vspace{1em}\hbox to 0.4\textwidth{\hfil
            $TOF(I_1;J_1;t_1) * TOF(I_2;J_2;t_2)$
        \hfil}\vspace{1.8em}}
        &
        \vbox{\vspace{1em}\hbox to 0.5\textwidth{\begin{minipage}{0.5\textwidth}\begin{center}
            $TOF(I_3;J_3;t_2) * TOF(I_1;J_1;t_1)$ \\
            $I_3 = (I_2 \setminus \{\,t_1\,\}) \cup (J_2 \cap \{\,t_1\,\})$
            $J_3 = (J_2 \setminus \{\,t_1\,\}) \cup (I_2 \cap \{\,t_1\,\})$
        \end{center}\end{minipage}}\vspace{0.5em}} \\

        \hline
        \multicolumn{2}{|c|}{\vbox{\vspace{0.4em}\hbox{
            Condition: gates are dependent, $I_1 \subseteq I_2$ and $J_1 \subseteq J_2$.
        }\vspace{0.05em}}} \\

        \hline
    \end{tabular}
\end{table}

So, the ``moving and replacing'' algorithm first searches a pair of gates,
the composition of which satisfies the condition of a replacement. After that the algorithm checks if they can be moved to
each other, using Lemma~\ref{lemma_gate_independency}. If yes, it implements a replacement as described above.
In the case, when the gate complexity is not reduced after replacement, but there are new gates in a circuit,
the algorithm continues to work, until the gate complexity is reduced or there are no new gates.

The time complexity of the proposed ``moving and replacing'' algorithm $T(A) \geq R \cdot l^2$, where $R$ is the number of
replacement rules, $l$ is the gate complexity of an original circuit. It is almost the same as the time complexity of any
template based optimization algorithm. At the same time, our ``moving and replacing'' algorithm seems to be more flexible
than a template-based approach, because proposed replacement rules do not depend on the number of inputs in a reversible circuit,
therefore there is no need to store a large number of templates and search them in a library.

\section{Boolean hypercube search}\label{section_cube}

Let's consider the following permutation:
\begin{align*}
    h = &   (\langle 1,0,0,0,0\rangle,\langle 1,0,1,0,1\rangle) \circ \\
    \circ & (\langle 1,0,0,0,1\rangle,\langle 1,0,1,0,0\rangle) \circ \\
    \circ & (\langle 1,0,0,1,0\rangle,\langle 1,0,1,1,1\rangle) \circ \\
    \circ & (\langle 1,0,0,1,1\rangle,\langle 1,0,1,1,0\rangle)  \; .
\end{align*}
As we can see, vectors in every transposition of permutation $h$ presented above differ only in the 3-rd and 5-th coordinates.
There are four transpositions total. Hence, a set of all vectors in these transpositions
represents a Boolean 3-cube $\mathbb B^{5,1,2}_{1,0}$ contained in a Boolean 5-cube $\mathbb B^5$. This 3-cube can also
be denoted as $\langle 1,0, *, *, * \rangle$.
Therefore, the permutation $h$ can be implemented by a composition of gates $TOF(1;2;3) * TOF(1;2;5)$.

Let's assume we can represent a permutation $h \in A(\BB^n)$ as a product of transpositions in such a way
that a set of all vectors of first $k$ transpositions in this product represents a Boolean $(1 + \log_2 k)$-cube.
In the case, when we use only our cycle-based approach for the synthesis,
we have to divide these $k$ transpositions into groups and
synthesize them separately. This approach can lead to significant gate complexity of a produced reversible circuit.
On the other hand, any Boolean hypercube contained in $\mathbb B^n$ can be implemented by a composition of no more than
$n$ generalized Toffoli gates $TOF(I;J;t)$.

For example, a transformation
$f\left ( \langle x_1, x_2, \cdots, x_n \rangle \right ) = \langle x_1, x_2 \oplus x_1, x_3, \ldots, x_n \rangle$
can be implemented by a reversible circuit, produced by our main synthesis algorithm, with the gate complexity $O(n2^n)$,
and it can be implemented by a single gate $TOF(1;2)$, because there is a Boolean 1-cube $\langle 1,*, \cdots, * \rangle$.

It is obvious that searching a Boolean hypercube can take a significant amount of time and can be inefficient
for large functions. But this approach makes it possible to obtain better synthesis results in some cases.

\subsection{Effective disjoints of cycles}

To find a larger Boolean hypercube, we should somehow effectively represent a permutation $h$ as a product of
specific transpositions. Let's consider a permutation $h = (a,b,c,e,f,g)$, where the Hamming distances
$d(a,e) = d(b,g) = d(c,f) = \mathrm\Delta$ and the Hamming distance for any other two elements of $h$
is not equal to $\mathrm\Delta$. We have the two possible representations of $h$ as a product of cycles:
\begin{enumerate}
    \item
        $h = (a,e)\circ(a,f,g)\circ(e,b,c)$.
    \item
        $h = (b,g)\circ(c,f)\circ(a,b)\circ(c,g)\circ(e,f)$.
\end{enumerate}
We can see that in the first case only the cycle $(a,e)$ has the two elements with the Hamming distance equal to $\mathrm\Delta$.
But in the second case there are two cycles $(b,g)$ and $(c,f)$ that have the two elements with the Hamming distance
equal to $\mathrm\Delta$.
Therefore, we can assume that the set $\{\,b,g,c,f\,\}$ may contain a larger Boolean hypercube,
compared to the set $\{\,a,e\,\}$, and we will call the second representation of $h$ an \textit{effective disjoint of cycles}.

There is a simple linear algorithm for an effective disjoint of cycles of a permutation for
a given Hamming distance $\mathrm\Delta$.
In the first pass, the algorithm searches all pairs of elements in a cycle with the Hamming
distance equal to $\mathrm\Delta$. In the second pass, the algorithm calculates for a found pair $p$
how many other pairs would be broken, if we disjoint the cycle by the pair $p$. In the third pass, the algorithm chooses
a pair $p$, for which the number of broken pairs is minimal. And finally, the algorithm disjoints the cycle by the chosen pair.
After that we don't have to repeat all steps for obtained cycles, because we can simply remove broken pairs and use previous
results for further disjoints.

For our example above, we have the three pairs with the Hamming distance equal to $\mathrm\Delta$: $(a,e)$, $(b,g)$ and $(c,f)$.
If we choose the pair $(a,e)$, it would break two pairs $(b,g)$ and $(c,f)$. And if we choose either $(b,g)$ or $(c,f)$,
they would break only the pair $(a,e)$. Hence, an effective disjoint will be for the pair $(b,g)$ or $(c,f)$.
It is not difficult to show that the proposed algorithm for an effective disjoint of cycles doesn't depend on the order of
elements in a cycle. The disjoint result will be the same for the permutation $h = (a,b,c,e,f,g)$ and for the
permutation $h' = (c,e,f,g,a,b)$.

The time complexity of a single disjoint operation for a cycle of length $l$ is no more than $O(l \log_2 l)$.

\subsection{Left and right multiplication}

Until now we used a left multiplication for a cycle disjoint. But we can also use a right multiplication.
E.\,g., a cycle $(a,b,c)$ can be represented in two ways for the transposition $(a,b)$:
\begin{enumerate}
    \item
        Left multiplication: $(a,b,c) = (a,b)\circ(a,c)$.
    \item
        Right multiplication: $(a,b,c) = (b,c)\circ(a,b)$.
\end{enumerate}
We can see that the results of the multiplications are different. This difference can lead to significantly different synthesis
results.

There is no way to find out on an $i$-th step of our basic synthesis algorithm,
whether the left or right multiplication would be the best in the end.
The only thing we can do is to make both left and right multiplications on an $i$-th step and choose the one
which leads to the greater permutation reduction and to the lower gate complexity of a current reversible circuit.
This approach doubles the synthesis time, but it also leads to better reversible circuits in some cases.

And finally, another area for optimizations is the constructing of a bijective transformation for a given non-bijective one.
We believe that in terms of reversible logic synthesis the best result can be achieved, when this bijective transformation
has minimal Hamming distances between inputs and outputs.

\section{Combining cycle-based and RM-spectra based algorithms}\label{section_rm_combination}

In~\cite{saeedi_cycle_based} a hybrid framework was proposed, which combines a cycle-based
and a RM-spectra based algorithms.
Unfortunately, this combination is only the choice of a better reversible circuit synthesized by one or another algorithm.

We propose a new approach for combining a cycle-based and a RM-spectra based algorithms.
In~\cite{maslov_rm_synthesis} a RM-spectra based synthesis algorithm was described.
For a reversible specification $f\colon \BB^n \to \BB^n$
the algorithm successively transforms the truth table $T_n$ to the truth table that corresponds to the identity transformation.
This is done by changing an $i$-th row in $T_n$, for which $T_n[i] \ne i$, to the form $T_n[i] = i$ for every
$i = 0, \cdots, (2^n - 1)$. Every row $j < i$ is not changed after a transformation of an $i$-th row: $T_n[j] = j$.

Our combining approach allows us to modify the truth table $T_n$ in such a way
that for the first row $i$, for which $T_n[i] \ne i$, $T_n[j] = j$, $j < i$, each row $k \leq i$
in a modified truth table $T'_n$ will be equal to itself: $T'_n[k] = k$.

On an $i$-th step of the RM-spectra based synthesis algorithm from~\cite{maslov_rm_synthesis}
a reversible circuit $\frS$ which we synthesize is of the form:
$$
    \frS = \frS_l * \frS_{T_n} * \frS_r \;  ,
$$
where sub-circuits $\frS_l$ and $\frS_r$ were constructed on the previous steps and a circuit $\frS_{T_n}$ is unknown, it
implements a transformation described by the truth table $T_n$, for which $T_n[j] = j$, $j < i$, $T_n[i] \ne i$.
The original RM-spectra based synthesis algorithm appends gates to the $\frS_l$ or $\frS_r$
after modifying the $i$-th row in $T_n$.

Let $h$, $h_l$, $h_t$, $h_r$ be the permutations,
defined by the circuits $\frS$, $\frS_l$, $\frS_{T_n}$ and $\frS_r$ respectively. This implies that
$$
    h = h_l \circ h_t \circ h_r  \; .
$$
Let's assume $T_n[i] = k$ and $T_n[l] = i$, where $k,l > i$. We can state that
\begin{align*}
    h =& h_l \circ h'_t \circ (i, k) \circ h_r = h_l \circ h'_t \circ h_r \circ (i, k)^{h_r} \;  ,\\
    h =& h_l \circ (i, l) \circ h'_t \circ h_r = (i,l)^{h_l^{-1}} \circ h_l \circ h'_t \circ h_r \;  ,
\end{align*}
where a permutation $h'_t$ is defined by the truth table $T'_n$, $T'_n[j] = j$ for every $j \leq i$.

From this it follows that we can ``push'' one transposition $(i,k)$ or $(i,l)$ from the permutation $h_t$
to the right or to the left, conjugate it by the permutation $h_r$ or $h_l^{-1}$ respectively
and ``skip'' the transformation of the $i$-th row in the truth table
$T_n$ by the original RM-spectra based synthesis algorithm. After that we can move to the next row and repeat this process.

After a RM-spectra based synthesis algorithm finishes its work, we can use a cycle-based synthesis algorithm
to synthesize pushed transpositions. There are several approaches to decide, whether an $i$-th row is pushed or not
and where it will be pushed (left or right). For example, we can push an $i$-th row only when the Hamming weight of $i$
is greater or equal to a predefined threshold $w$. It is equivalent to processing all monomials of degree $d < w$
in a Reed--Muller polynomial with a RM-spectra based synthesis algorithm. All other monomials will be processed
by a cycle-based synthesis algorithm.

\begin{figure}
    \centerline{
        \Qcircuit @C=1em @R=.7em @!R {
            \lstick{x_1} & \qw      & \qw        & \targ     & \ctrlo{1} & \ctrl{3}  & \ctrl{3} & \qw       & \ctrl{1}  & \ctrlo{1} & \qw      & \qw      & \rstick{*}   \qw \\
            \lstick{x_2} & \ctrl{1} & \targ      & \ctrl{-1} & \ctrl{4}  & \qw       & \qw      & \ctrlo{1} & \ctrlo{1} & \ctrl{2}  & \qw      & \qw      & \rstick{*}   \qw \\
            \lstick{x_3} & \ctrl{3} & \ctrl{-1}  & \qw       & \qw       & \qw       & \qw      & \ctrl{1}  & \ctrl{1}  & \qw       & \qw      & \qw      & \rstick{*}   \qw \\
            \lstick{x_4} & \qw      & \qw        & \qw       & \qw       & \ctrl{2}  & \targ    & \ctrl{1}  & \ctrlo{3} & \ctrl{1}  & \ctrl{1} & \ctrl{1} & \rstick{*}   \qw \\    
            \lstick{x_5} & \qw      & \qw        & \qw       & \qw       & \qw       & \qw      & \ctrl{2}  & \qw       & \ctrl{2}  & \ctrl{1} & \targ    & \rstick{y_1} \qw \\
            \lstick{0}   & \targ    & \qw        & \qw       & \targ     & \targ     & \qw      & \qw       & \qw       & \qw       & \targ    & \qw      & \rstick{y_2} \qw \\
            \lstick{0}   & \qw      & \qw        & \qw       & \qw       & \qw       & \qw      & \targ     & \targ     & \targ     & \qw      & \qw      & \rstick{y_3} \qw 
        }
    }
    \caption{Realization of \textbf{rd53} function in a reversible circuit with 7 inputs and 11 gates in the GT library.}
    \label{fig_rd53}
\end{figure}
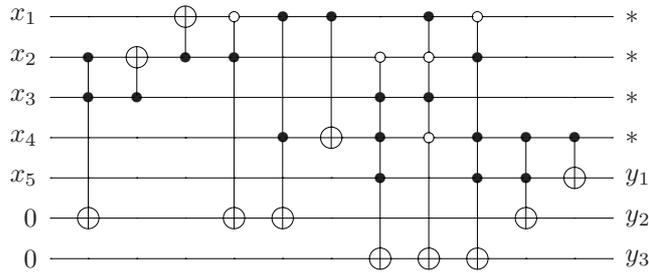

We realized the proposed combining approach in our software~\cite{my_software}
with the ability to choose a ``push policy'' and a weight threshold.
This allowed us to find a reversible circuit implementing \textbf{rd53} function
with 7 inputs and with the gate complexity equal to 11 in the GT library (see Fig.~\ref{fig_rd53}).
The weight threshold was equal to one during the synthesis process.

\section{Experimental results}\label{section_exp}

We developed an open source software \textit{ReversibleLogicGenerator}~\cite{my_software},
which implements the basic cycle-based synthesis algorithm from~\cite{my_gate_complexity}
and all the gate complexity reduction techniques, described in this paper.
Using our software, we conducted a series of experiments on reversible benchmark functions synthesis. The results are
presented in Tables~\ref{table_less_inputs}--\ref{table_less_cost}.
The synthesis time in the worst case was a matter of seconds.

\setlength\tabcolsep{2pt}
\begin{table}
    \renewcommand{\arraystretch}{1.3}
    \caption{Benchmark functions synthesis (new circuits with less input count).}
    \label{table_less_inputs}
    \centering
    \begin{threeparttable}
        \small
    \begin{tabular}{|c||c|c|c|c||c|c|c|c|}
        \hline
        
        \multirow{2}{*}{
            \textbf{Function}
        } &
        \multicolumn{4}{c||}{\textbf{New circuits}} &
        \multicolumn{4}{c|}{\textbf{Existing circuits}} \\
        \cline{2-9}
        
        & lines$_{\min}$ & GC & QC & $T$-count & lines$_{\min}$ & GC & QC & $T$-count\\
        \hline
        
        gf2\verb"^"3mult & \textbf{7} & 73 & 740 & 632 &
            \multirow{3}{*}{\textbf{9}} & \multirow{3}{*}{11} & \multirow{3}{*}{47} & \multirow{3}{*}{63} \\
        \cline{1-5}
        gf2\verb"^"3mult & \textbf{7} & 79 & 712 & 632 & & & & \\
        \cline{1-5}
        gf2\verb"^"3mult & \textbf{7} & 145 & 704 & 654 & & & & \\
        \hline
        gf2\verb"^"4mult & \textbf{9} & 415 & 47649 & 10838\tnote{*} &
            \multirow{2}{*}{\textbf{12}} & \multirow{2}{*}{19} & \multirow{2}{*}{83} & \multirow{2}{*}{112} \\
        \cline{1-5}
        gf2\verb"^"4mult & \textbf{9} & 1834 & 5914 & 5156 & & & & \\
        \hline
        nth\_prime9\_inc & \textbf{9} & 3942 & 19313 & 15234 & \textbf{10} & 7522 & 17975 & 14193 \\
        \hline
        rd73 & \textbf{9} & 296 & 43421 & 8765\tnote{*} &
            \multirow{2}{*}{\textbf{10}} & \multirow{2}{*}{20} & \multirow{2}{*}{64} & \multirow{2}{*}{98} \\
        \cline{1-5}
        rd73 & \textbf{9} & 835 & 4069 & 3521 & & & & \\
        \hline            
        rd84 & \textbf{11} & 679 & 359384 & 25364\tnote{*} &
            \multirow{2}{*}{\textbf{15}} & \multirow{2}{*}{28} & \multirow{2}{*}{98} & \multirow{2}{*}{147} \\
        \cline{1-5}
        rd84 & \textbf{11} & 2560 & 12397 & 8772\tnote{*} & & & & \\
        \hline
    \end{tabular}
    \begin{tablenotes}
        \item[*] an ancillary line is required
    \end{tablenotes}
    \end{threeparttable}
\end{table}

\begin{table}
    \renewcommand{\arraystretch}{1.3}
    \caption{Benchmark functions synthesis (new circuits with less gate complexity).}
    \label{table_less_complexity}
    \centering
    \begin{threeparttable}
    \begin{tabular}{|c||c|c|c|c||c|c|c|c|}
        \hline
        
        \multirow{2}{*}{
            \textbf{Function}
        } &
        \multicolumn{4}{c||}{\textbf{New circuits}} &
        \multicolumn{4}{c|}{\textbf{Existing circuits}} \\
        \cline{2-9}
        
        & lines & GC$_{\min}$ & QC & $T$-count & lines & GC$_{\min}$ & QC & $T$-count\\
        \hline
        
        2of5 & 6 & \textbf{9} & 268 & 191\tnote{*} &
            \multirow{2}{*}{6} & \multirow{2}{*}{\textbf{15}} & \multirow{2}{*}{107} & \multirow{2}{*}{119} \\
        \cline{1-5}
        2of5 & 6 & \textbf{10} & 118 & 135 & & & & \\
        \hline
        2of5 & 7 & \textbf{11} & 32 & 42 & 7 & \textbf{12} & 32 & 49 \\
        \hline
        3\_17 & 3 & \textbf{4} & 14 & 14 &
            \multirow{2}{*}{3} & \multirow{2}{*}{\textbf{6}} & \multirow{2}{*}{12} & \multirow{2}{*}{14} \\
        \cline{1-5}
        3\_17 & 3 & \textbf{5} & 13 & 14 & & & & \\
        \hline
        4b15g\_2 & 4 & \textbf{12} & 57 & 55\tnote{*} & 4 & \textbf{15} & 31 & 35 \\
        \hline
        4b15g\_4 & 4 & \textbf{12} & 49 & 45\tnote{*} &
            \multirow{2}{*}{4} & \multirow{2}{*}{\textbf{15}} & \multirow{2}{*}{35} & \multirow{2}{*}{31\tnote{*}} \\
        \cline{1-5}
        4b15g\_4 & 4 & \textbf{14} & 47 & 45\tnote{*} & & & & \\
        \hline
        4b15g\_5 & 4 & \textbf{14} & 72 & 54\tnote{*} & 4 & \textbf{15} & 29 & 42 \\
        \hline
        4mod5 & 5 & \textbf{4} & 13 & 14 & 5 & \textbf{5} & 7 & 7 \\
        \hline
        5mod5 & 6 & \textbf{7} & 429 & 294\tnote{*} & 6 & \textbf{8} & 84 & 70\tnote{*} \\
        \hline
        6sym & 7 & \textbf{14} & 1308 & 628\tnote{*} &
            \multirow{2}{*}{7} & \multirow{2}{*}{\textbf{36}} & \multirow{2}{*}{777} & \multirow{2}{*}{741} \\
        \cline{1-5}
        6sym & 7 & \textbf{15} & 825 & 624\tnote{*} & & & & \\
        \hline
        9sym & 10 & \textbf{73} & 61928 & 7004\tnote{*} &
            \multirow{2}{*}{10} & \multirow{2}{*}{\textbf{129}} & \multirow{2}{*}{6941} & \multirow{2}{*}{5484\tnote{*}} \\
        \cline{1-5}
        9sym & 10 & \textbf{74} & 31819 & 6788\tnote{*} & & & & \\
        \hline
        ham7 & 7 & \textbf{19} & 77 & 85 & 7 & \textbf{25} & 49 & 42 \\
        \hline
        hwb12 & 12 & \textbf{42095} & 134316 & 98482 & 12 & \textbf{55998} & 198928 & 134131 \\
        \hline
        nth\_prime7\_inc & 7 & \textbf{427} & 10970 & 5403\tnote{*} &
            \multirow{3}{*}{7} & \multirow{3}{*}{\textbf{1427}} & \multirow{3}{*}{3172} & \multirow{3}{*}{2837} \\
        \cline{1-5}
        nth\_prime7\_inc & 7 & \textbf{474} & 10879 & 5403\tnote{*} & & & & \\
        \cline{1-5}
        nth\_prime7\_inc & 7 & \textbf{824} & 2269 & 1906 & & & & \\
        \hline
        nth\_prime8\_inc & 8 & \textbf{977} & 10218 & 7359\tnote{*} &
            \multirow{2}{*}{8} & \multirow{2}{*}{\textbf{3346}} & \multirow{2}{*}{7618} & \multirow{2}{*}{5985} \\
        \cline{1-5}
        nth\_prime8\_inc & 8 & \textbf{1683} & 6330 & 5213 & & & & \\
        \hline
        nth\_prime9\_inc & 10 & \textbf{2234} & 22181 & 17292 & 10 & \textbf{7522} & 17975 & 14193 \\
        \hline
        nth\_prime10\_inc & 11 & \textbf{5207} & 50152 & 38261 & 11 & \textbf{16626} & 40299 & 30315 \\
        \hline
        nth\_prime11\_inc & 12 & \textbf{11765} & 124408 & 92937 & 12 & \textbf{35335} & 95431 & 68255 \\
        \hline
        rd53 & 7 & \textbf{11} & 96 & 100 & 7 & \textbf{12} & 120 & 124 \\
        \hline
    \end{tabular}
    \begin{tablenotes}
        \item[*] an ancillary line is required
    \end{tablenotes}
    \end{threeparttable}
\end{table}

\begin{table}
    \renewcommand{\arraystretch}{1.3}
    \caption{Benchmark functions synthesis (new circuits with less quantum cost).}
    \label{table_less_cost}
    \centering
    \begin{threeparttable}
    \begin{tabular}{|c||c|c|c|c||c|c|c|c|}
        \hline
        
        \multirow{2}{*}{
            \textbf{Function}
        } &
        \multicolumn{4}{c||}{\textbf{New circuits}} &
        \multicolumn{4}{c|}{\textbf{Existing circuits}} \\
        \cline{2-9}
        
        & lines & GC & QC$_{\min}$ & $T$-count & lines & GC & QC$_{\min}$ & $T$-count\\
        \hline
        
        2of5 & 7 & 12 & \textbf{31} & 42 & 7 & 12 & \textbf{32} & 49 \\
        \hline
        6sym & 7 & 41 & \textbf{206} & 184 & 7 & 36 & \textbf{777} & 741 \\
        \hline
        9sym & 10 & 347 & \textbf{1975} & 1680 & 10 & 210 & \textbf{4368} & 4368 \\
        \hline
        hwb7 & 7 & 603 & \textbf{1728} & 1400 & 7 & 331 & \textbf{2611} & 2245\tnote{*} \\
        \hline
        hwb8 & 8 & 1594 & \textbf{4852} & 3748 & 8 & 2710 & \textbf{6940} & 5201 \\
        \hline
        hwb9 & 9 & 3999 & \textbf{12278} & 10220 & 9 & 6563 & \textbf{16173} & 12150 \\
        \hline
        hwb10 & 10 & 8247 & \textbf{26084} & 20368 & 10 & 12288 & \textbf{35618} & 25939 \\
        \hline
        hwb11 & 11 & 21432 & \textbf{69138} & 52922 & 11 & 32261 & \textbf{90745} & 63430 \\
        \hline
        hwb12 & 12 & 42095 & \textbf{134316} & 98482 & 12 & 55998 & \textbf{198928} & 134131 \\
        \hline
        nth\_prime7\_inc & 7 & 824 & \textbf{2269} & 1906 & 7 & 1427 & \textbf{3172} & 2837 \\
        \hline
        nth\_prime8\_inc & 8 & 1683 & \textbf{6330} & 5213 & 8 & 3346 & \textbf{7618} & 5985 \\
        \hline
        rd53 & 7 & 12 & \textbf{82} & 92 &
            \multirow{3}{*}{7} & \multirow{3}{*}{12} & \multirow{3}{*}{\textbf{120}} & \multirow{3}{*}{124} \\
        \cline{1-5}
        rd53 & 7 & 12 & \textbf{95} & 100 & & & & \\
        \cline{1-5}
        rd53 & 7 & 11 & \textbf{96} & 100 & & & & \\
        \hline
    \end{tabular}
    \begin{tablenotes}
        \item[*] an ancillary line is required
    \end{tablenotes}
    \end{threeparttable}
\end{table}

We were able to obtain more than 40 new reversible circuits, which have less input count, less gate complexity or
less quantum cost compared to existing circuits.
All specifications for benchmark functions and their names were taken from the
\textit{Reversible Logic Synthesis Benchmarks Page}~\cite{maslov_benchmark} and from the \textit{RevLib} site~\cite{revlib}.
We use the following conventions in the tables: \textit{lines} is the number of inputs in a reversible circuit $\frS$,
GC is the gate complexity $L(\frS)$ of this circuit and QC is its quantum cost $W(\frS)$;
$T$-count is the number of $T$ gates in a decomposition of the circuit into Clifford$+T$ gates.

The quantum cost of obtained circuits was calculated with the help of the software \textit{RCViewer+}~\cite{rcviewer}.
Its calculation is based on the paper~\cite{quantum_cost},
according to which a generalized Toffoli gate with negative control lines
may have the same quantum cost as the corresponding generalized Toffoli gate without negative control lines.
Also we included the $T$-count cost measure for all circuits in the tables
(cost calculation was based on the paper~\cite{maslov_t_count}). Despite the fact that this cost measure is very
popular for fault-tolerant circuits in the literature, it is not universal in the case of limited ancillary lines availability.
According to~\cite{maslov_t_count}, there is a circuit of Toffoli gates that cannot be implemented, using Clifford$+T$ gates,
without an ancillary line. We marked such $T$-count metrics by the asterisk symbol in the tables.

Tables~\ref{table_less_inputs}, \ref{table_less_complexity} and \ref{table_less_cost} contain results for obtained
reversible circuits with less input count (column lines$_{\min}$), less gate complexity (column GC$_{\min}$)
and less quantum cost (column QC$_{\min}$) compared to existing circuits respectively.
Since we compare our circuits only with circuits consisting of gates from NCT and GT libraries
and since the NCT library is a part of the GT library, such comparison made by us is correct.

We have not included circuits with more than 12 inputs in the tables just because of the limited format of the paper.
One can easily synthesize such circuits with the help of our software.

With the help of developed software we were able to find a reversible circuit with 7 input lines and with 11 gates from the GT
library for one of the most popular benchmark functions \textbf{rd53} (see Fig.~\ref{fig_rd53}).
This circuit and all other circuits described in the tables above can be freely downloaded in TFC and REAL formats from
the cite~\cite{my_software} as well as \textit{ReversibleLogicGenerator} software itself.

\section{Conclusion}

A reversible circuit with $n$ inputs necessarily defines a permutation from the symmetric group $S(\BB^n)$.
Permutations that correspond to all the gates NOT, CNOT and C\textsuperscript{2}NOT generate the alternating group $A(\BB^n)$
if $n > 3$, and permutations that correspond to all the gates C\textsuperscript{k}NOT,
$1 \leq k \leq n$ generate the symmetric group.
This implies that we can use the permutation group theory to successfully synthesize a reversible circuit for a given
reversible specification.

In the paper, we briefly described the first asymptotically optimal in NCT library synthesis algorithm, based on the permutation
group theory, which makes it possible to obtain a reversible circuit without additional inputs.
We also suggested the ``moving and replacing'' algorithm for gate complexity reduction for circuits consisting of the
gates from the GT library; the algorithm is based on equivalent replacements of gate compositions
and on conditions of independence for the gates with negative control lines.

We described some gate complexity reduction techniques that use the permutation group theory. Among them are
the search of a Boolean hypercube and an effective cycle disjoint.
We presented experimental results for benchmark functions synthesis, which include more than 40 reversible circuits consisting
of gates from the GT library, obtained with the help of developed open source software that implements all described techniques.

We believe that the permutation group theory may allow us to obtain better reversible circuits for all benchmark functions,
and we hope that this paper will motivate other researchers to improve our results.

\subsubsection*{Acknowledgments.} The reported study was partially supported by RFBR, research project No. 16-01-00196~A.

\bibliographystyle{abbrv}
\bibliography{IEEEabrv,references}

\end{document}